# Deep Learning to Quantify Pulmonary Edema in Chest Radiographs

**Steven Horng, MD, MMSc[1,2]\*, Ruizhi Liao[3]\*, Xin Wang, PhD[4], Sandeep Dalal[4], Polina Golland, PhD[3], Seth J Berkowitz, MD[1,2]**

1: Beth Israel Deaconess Medical Center, Boston, MA

2: Harvard Medical School, Boston, MA

3: Massachusetts Institute of Technology, Cambridge, MA

4: Philips Research, Cambridge, MA

**\* The two first authors contributed equally**

# ABSTRACT

**Purpose:**

To develop a machine learning model to classify the severity grades of pulmonary edema on chest radiographs.

**Materials and Methods:**

In this retrospective study, 369,071 chest radiographs and associated radiology reports from 64,581 (mean age, 51.71; 54.51% women) patients from the MIMIC-CXR chest radiograph dataset were included. This dataset was split into patients with and without congestive heart failure (CHF). Pulmonary edema severity labels from the associated radiology reports were extracted from patients with CHF as four different ordinal levels: 0, no edema; 1, vascular congestion; 2, interstitial edema; and 3, alveolar edema. Deep learning models were developed using two approaches: a semi-supervised model using a variational autoencoder and a pre-trained supervised learning model using a dense neural network. Receiver operating characteristic curve analysis was performed on both models.

**Results:**

The area under the receiver operating characteristic curve (AUC) for differentiating alveolar edema from no edema was 0.99 for the semi-supervised model and 0.87 for the pre-trained models. Performance of the algorithm was inversely related to the difficulty in categorizing milder states of pulmonary edema (shown as AUCs for semi-supervised model and pre-trained model, respectively): 2 versus 0, 0.88 and 0.81; 1 versus 0, 0.79 and 0.66; 3 versus 1, 0.93 and 0.82; 2 versus 1, 0.69 and 0.73; and, 3 versus 2, 0.88 and 0.63.

**Conclusion:**

Deep learning models were trained on a large chest radiograph dataset and could grade the severity of pulmonary edema on chest radiographs with high performance.

# 1. INTRODUCTION

Chest radiographs are commonly performed to assess pulmonary edema (1). The signs of pulmonary edema on chest radiographs have been known for over 50 years (2,3). The grading of pulmonary edema is based on well-known radiologic findings on chest radiographs [4–7]. The symptom of dyspnea caused by pulmonary edema is the most common reason a patient with acute decompensated congestive heart failure (CHF) seeks care in the emergency department and is ultimately admitted to the hospital (89% of patients) (8–10). Clinical management decisions for patients with acutely decompensated CHF are often based on grades of pulmonary edema severity, rather than its mere absence or presence. Clinicians often monitor changes in pulmonary edema severity to assess the efficacy of therapy. Accurate monitoring of pulmonary edema is essential when competing clinical priorities complicate clinical management (additional information in **Appendix E1 [supplement]**).

While we focus on patients with CHF within this study, the quantification of pulmonary edema on chest radiographs is useful throughout clinical medicine. Pulmonary edema is a manifestation of volume status in sepsis and renal failure, just as in CHF. Managing volume status is critical in the treatment of sepsis, but large-scale research has been limited due to longitudinal data on volume status. Quantification of pulmonary edema in a chest radiograph could be used as a surrogate for volume status, which would rapidly advance research in sepsis and other disease processes where volume status is critical.

Large-scale and common datasets have been the catalyst for the rise of machine learning today (11). In 2019, investigators released MIMIC-CXR, a large-scale publicly available chest radiograph dataset (12–15). This investigation builds upon that prior work by developing a common, clinically meaningful machine learning task and evaluation framework with baseline performance metrics to benchmark future algorithmic developments in grading pulmonary edema severity from chest radiographs. We developed image models using two common machine learning approaches: a semi-supervised learning model and a supervised learning model pre-trained on a large common image dataset.

## 2. MATERIALS AND METHODS

**2.1 Study Design**

This was a retrospective cohort study. This study was approved by the Beth Israel Deaconess Medical Center Committee on Clinical Investigation with a waiver of informed consent. We collected 369,071 chest radiographs and their associated radiology reports from 64,581 patients from the MIMIC-CXR chest radiograph dataset (12–14). Each imaging study is associated with one or more images. We aimed to identify patients with CHF within the dataset to limit confounding labels from other disease processes. First, we limited our study to only frontal radiographs, excluding a total of 121 646 images. Of these frontal radiographs ($n = 247\,425$), there were 17,857 images which were acquired during visits with an emergency department discharge diagnosis code consistent with CHF. In total, this resulted in 16,108 radiology reports and 1,916 patients that were included that had CHF. As part of a prior study (26), we manually reviewed patient charts and found this method of cohorting patients with CHF had 100% sensitivity and specificity. The other 62 665 patients were classified as non-CHF and data was used in the semi-supervised training model. An enrollment diagram is shown in **Figure 1**.

**2.2 Label Extraction and Validation**

We extracted the pulmonary edema severity labels ("none", "vascular congestion", "interstitial edema", and "alveolar edema") from the reports using regular expressions with negation detection. The extracted labels were numerically coded as follows: 0, none; 1, vascular congestion; 2, interstitial edema; and 3, alveolar edema **(Table 1)**. Examples of the grades are shown in **Figure E1** (**supplement**). We were able to label 3,028 radiology reports and thus 3,354 frontal view radiographs from 1,266 patients (**Figure 1**). Among the 1,266 patients, 1,180 patients still have some of their reports unlabeled. The other 650 patients with CHF had no labeled reports.

To validate our label extraction in radiology reports, we randomly selected 200 labeled reports (50 for each severity category from patients with CHF). A board-certified radiologist (SB, 5 years of experience, interventional radiology) then manually labeled the 200 reports, blinded from our label extraction results. We report the precision (positive predictive value) of the regular expression results for each category and each keyword, and sensitivity and specificity of each keyword.

We had three senior radiology residents and one attending radiologist manually label a set of 141 frontal view radiographs from 123 patients (from the unlabeled dataset of 650 patients with CHF), which had no patient overlap with the report labeled set (**Figure E2** [**supplement**]). These images were set aside as our test set. Each radiologist assessed the images independently and we report their inter-rater agreement (Fleiss' Kappa). We used a modified Delphi consensus process, further described in **Appendix E1** (**supplement**), to develop a consensus reference standard label.

## 2.3 Model Development

In order to establish a baseline performance benchmark for this clinical machine learning task and to address the challenge of limited pulmonary edema labels, we developed models using two common computer vision approaches: a semi-supervised model using a variational autoencoder (16) and a pre-trained supervised learning model using a dense neural network (17,18). Both approaches aim to address the challenge of limited pulmonary edema labels. The first approach

(semi-supervised model) takes advantage of the chest radiographs without pulmonary edema severity labels, which includes approximately 220,000 images (from individuals with and without CHF) and is domain specific. The second approach (pre-trained supervised model) uses a large-scale common image dataset with common object labels (such as cats and dogs), which includes approximately 14M images and leverages the image recognition capability from other domains.

In order to mitigate the imbalanced dataset size of each severity level, we employ weighted cross entropy as the loss term for training both models. Data augmentation (including random translation and rotation) is performed during training to accommodate the variable patient positionings.

*Semi-supervised Learning Model Development.*

To take advantage of the large number of unlabeled chest radiographs, we developed a Bayesian model that includes a variational autoencoder for learning a latent representation from the entire radiograph set (exclusive of the test set) trained jointly with a classifier that employs this representation for estimating edema severity. We first trained the variational autoencoder on both unlabeled and labeled images (exclusive of the test set), although the labels were not involved at this stage. The variational autoencoder learned to encode the chest radiographs into compact (low-dimensional) image feature representations by an encoder, and learned to reconstruct the images from the feature representation by a decoder. We then took the trained encoder and concatenated it with an image classifier that estimates pulmonary edema severity. Finally, we trained this encoder with the classifier on labeled images in a supervised learning fashion. The use of this variational autoencoder architecture allowed us to leverage a large number of unlabeled images to train a model that learns the underlying features of chest radiograph images. By training the variational autoencoder jointly with a classifier on the labeled images, we ensure it captures compact feature representations for scoring pulmonary edema severity. We also use data augmentation by random image translation, rotation, and cropping to a size of 2048 x 2048 during training in order to improve the robustness of the model. We use deep convolutional neural networks to implement the variational autoencoder and the classifier. The encoder of the variational autoencoder has eight residual blocks (5), the decoder has five deconvolution layers, and the classifier has four residual blocks followed by two fully-connected layers.

We also varied the number of unlabeled chest radiographs used to train this semi-supervised model, in order to assess how the model performance changes with the amount of unlabeled data. We report the average of the nine area under the receiver operating characteristic curve (AUC) values (as in **Table 4**) in the **Table E1** (**supplement**).

*Pretrained Model Development.*

In the second approach, we started with a neural network that had been pre-trained to recognize common images (e.g., cats and dogs) and then further tuned it to recognize the specific image features of chest radiographs for assessing pulmonary edema. Specifically, we use the densely connected convolutional neural networks (DenseNet) (6) and the model is pre-trained on ImageNet (7). The DenseNet has four dense blocks (6), which consist of 6, 12, 24, 16 convolutional layers respectively. The four dense blocks are concatenated with a 2-by-2 averaging pooling layer between each two consecutive dense blocks. We keep the first three pre-trained dense blocks for low-level image feature extraction, followed by one global average pooling layer, one dropout layer and two fully connected layers. We then re-trained this model on our labeled chest radiographs. We also use data augmentation by random image translation, rotation, and cropping to a size of 512 x 512 (for adjusting the image size in the ImageNet) during training in order to improve the robustness of the model.

**2.4 Statistical Analysis**

Study population means and 95% CIs were reported for age, and percentages were reported for sex and disposition. A Student's t-test was used to test for significance for age, and a Pearson chi-squared test was used for sex and disposition.

To understand how many and how frequently chest radiographs have been taken on our CHF cohort and non-CHF cohort, we calculated the number of images from each patient in our dataset and plotted the histograms of the numbers for the CHF cohort and for the non-CHF cohort. We also showed the distributions of time intervals between two consecutive chest radiographs taken on a patient with CHF.

To evaluate the model, we performed five-fold cross-validation and randomly split the 3,354 labeled images into five folds, ensuring that images from the same patients were allocated to the same fold. For each round, four folds were used for training and the remaining fold was held out for evaluation. Each model was trained five times independently to evaluate all five folds. During training, the validation fold was never seen by the model. We selected the best trained model among the five and tested it on the manually labeled image test set. The distribution of severity labels across folds and the test set is summarized in **Table 2**. The cross-validation results are summarized in **Appendix E1** (**supplement**).

We plotted receiver operating characteristic curves (ROC) and reported the AUC for each pairwise comparison between severity labels on the test set. We then dichotomized the severity and reported three comparisons: *(a)* 0 versus 1,2,3; *(b)* 0,1 versus 2,3; and *(c)* 0,1,2 versus 3. We used the DeLong method to test for significance between AUC's between the semi-supervised model and the pre-trained model. In order to account for multiple comparisons, a Bonferroni correction was used with $\alpha = 0.05 \div 9 = 0.005$.

Lastly, we show the confusion matrices for each of the models. To interpret the model predictions, we use Grad-CAM to produce heatmaps to visualize the areas of the radiographs that are most informative for grading pulmonary edema severity. Grad-CAM computes the gradients of the model prediction with respect to the feature maps of the last convolutional layer in the model. The gradients are used to calculate the weighted average of the feature maps and the weighted average map is displayed as a heatmap to visualize image regions that are "important" for the model prediction (19).

**2.7 Data Availability**

The underlying data is available at https://github.com/RayRuizhiLiao/mimic_cxr_edema.

# 3. RESULTS

**Patient and Chest Radiograph Characteristics**

We analyzed the chest radiograph distributions in our CHF cohort (1,916 patients) and non-CHF cohort (62,665 patients). The histograms for number of chest radiographs and interval time is shown in **Figure E3** (**supplement**). The mean number of chest radiographs taken per patient with CHF was 14 (median, 9; range 1-153) and per patient with no CHF was 5 (median, 3; range 1-174). For patients with CHF, the mean interval time between each two consecutive chest radiograph orders from the same patient was 71 days (median, 7 days; range 0.13-1545). A total of 21.53% of patients had interval times within 1 day, while 66.08% had interval times within 30 days. Additional information on radiographs and patients are shown in **Table 3**.

**Validation of Outcome Measures**

The precision values (positive predictive value) of the regular expression results (i.e., extracting pulmonary edema severity labels from the radiology reports within the dataset) for "none", "vascular congestion", "interstitial edema", and "alveolar edema" based on the manual review results were 96%, 84%, 94%, and 94%, respectively. The overall precision was 92%. The precision, sensitivity, and specificity for each keyword are summarized in **Table 1**.

After independent labeling, discussion, and voting, the inter-rater agreement (Fleiss' Kappa) among the three radiology residents was 0.97 (more details in **Figure E2** [**supplement**]). Our modified Delphi process yields consensus labels for all 141 images.

**Receiver Operating Characteristics Curve Analysis**

The ROC curves of the two models on the test set are shown in (**Figure 2**). As expected, both models perform well on the task of distinguishing images between level 0 and level 3 and on the task of classifying between level 3 and the rest. The AUC for differentiating alveolar edema (score 3) from no edema (score 0) was 0.99 and 0.87 for semi-supervised and pre-trained models, respectively. Performance of the algorithm was inversely related to the difficulty in categorizing milder states of pulmonary edema (shown as the AUC for the semi-supervised and pretrained model, respectively, for differentiating the following categories): 2 versus 0, 0.88 and 0.81; 1 versus 0, 0.79 and 0.66; 3 versus 1, 0.93 and 0.82; 2 versus 1, 0.69 and 0.73); 3 versus 2, 0.88 and 0.63. The ROC curves from the cross-validation are shown in **Figure E4** (**supplement**).

The AUCs of the two models on the test set are reported in **Table 4**. Seven out of the nine Delong test significance values were higher than .005, which means that the two models did not have significantly different AUCs. The AUCs of the cross-validation results are reported in **Table E2** (**supplement**).

**Confusion Matrix Analysis**

We computed a confusion matrix for each of the models on the test set (**Figure 3**). Each image was placed in a cell by the true severity level from consensus score and the predicted severity level from the image model. In each cell, we reported the fraction of the predicted severity level in the actual severity level. Both models performed better in predicting level 0 and level 3 compared to predicting level 1 and level 2. The confusion matrices from the cross-validation are summarized in **Figure E5** (**supplement**).

**Predicted Edema Severity in Bar Charts**

We plotted bar charts of predicted edema severity versus true edema severity on the test set (**Figure 4**). Both plots show the linear trend of predicted edema severity with ground truth edema severity. Overlap of error bars graphically depicts the challenges in discriminating less severe stages of pulmonary edema. Pulmonary edema severity exists on a continuous spectrum and future work on this will be discussed in the following section. Similar bar charts from the cross-validation are reported in **Figure E6** (**supplement**).

**Model Interpretation**

We used Grad-CAM to visualize the regions in a radiograph that are important for the model prediction. (**Figure 5)** demonstrates two sample images from the two models. We also manually reviewed the test data set in an attempt to classify the failure modes of both the semi-supervised and pre-trained models (**Table E3** [**supplement**]).

## 4. DISCUSSION

We have employed two different machine learning techniques to quantify pulmonary edema. The semi-supervised approach learns from all the radiographs in the training set. The pre-trained image model learns from a large common image set and the labeled radiographs. Both approaches aim to address the challenge of limited labels extracted from the radiology reports. Both approaches have similar performance statistically in terms of AUC on most pairwise classification comparisons (seven out of nine). On the other two comparisons (two out of nine), the semi-supervised approach outperforms the pre-trained approach. The semi-supervised approach may give better results because it has learned from approximately 220,000 chest radiographs and is thus tailored to the image feature extraction of chest radiographs.

The semi-supervised model was rarely off by two levels of pulmonary edema and never disagreed by three levels from the consensus label. However, there were examples where the pretrained model predicted alveolar edema or no pulmonary edema when the consensus label was on the opposite end of the spectrum. More work is needed to improve the explainability of the model to understand these failure modes which are clearly critical before such a model could be deployed in clinical practice. Importantly, however, the manual review showed several examples where the models were able to correctly assess the absence of pulmonary edema despite the presence of severe cardiomegaly and pleural effusions.

The results of these algorithms provide a performance benchmark for future work. We have shown that it is feasible to automatically classify four levels of pulmonary edema on chest radiographs. Understandably, the performance of the algorithm mirrors the challenge of distinguishing these disease states for radiologists. The differentiation of alveolar edema from no pulmonary edema (level 3 vs 0) is an easier task than distinguishing interstitial edema from pulmonary vascular congestion (level 2 vs 1). Even among radiologists, there is substantial variability in the assessment of pulmonary edema. More machine learning approaches should be explored for this clinical task in future work.

Our work expands on prior studies by employing machine learning algorithms to automatically and quantitatively assess the severity of pulmonary edema from chest radiographs. Prior work has shown the ability of convolutional neural networks to detect pulmonary edema among several other pathologies that may be visualized in chest radiographs (20–22). Neural networks have been validated in large datasets to achieve expert level identification of findings in chest radiographs (23). Their AUCs in detecting the presence of pulmonary edema range from 0.83 to 0.88. By treating pulmonary edema as a single pathology, it is difficult to draw direct comparison to our work which considers pulmonary edema as a spectrum of findings. A conservative comparison would be to compare prior work to our model's ability to distinguish no edema and pulmonary vascular congestion from interstitial and alveolar edema (levels 0,1 vs 2,3) which have AUCs of 0.81 (pre-trained) and 0.88 (semi-supervised). Although their test sets are based on labels extracted from radiology reports, our test set labels are annotated and reached consensus on by four radiologists. Others have trained neural networks on B-type natriuretic peptide values to produce a quantitative assessment of congestive heart failure (24). However, B-type natriuretic peptide increases non-linearly with worsening CHF, and exhibits marked inter-patient variability. A B-type natriuretic peptide of 1000 in one patient could represent an acute exacerbation, while being the baseline for another patient, making B-type natriuretic peptide a poor surrogate outcome measure for acute pulmonary edema. The grading of pulmonary edema severity relies on much more subtle radiological findings (image features). The clinical management of patients with pulmonary edema requires comparisons of serial exams and understanding serial trends. Accurate, reproducible, and rapid quantification of pulmonary edema is of paramount value to clinicians caring for these patients.

There were limitations in our study. Extracting labels from clinical radiology reports allowed us to quickly obtain a reasonable amount of labelled data, but is inferior to data labelled for a specific purpose. Not only is there poor inter-reader agreement among radiologists for pulmonary edema detection (25), but radiologists may use different languages to describe a similar pathophysiologic state. In future work, we will explore joint modeling of chest radiographs and radiology reports and aim to mitigate the bias introduced by simply employing regular expressions.

Pulmonary edema exists on a continuous spectrum of severity. By discretizing our data into four classes, we have potentially lost valuable information and contaminated the categories. The category of severe edema in our dataset contains all images containing alveolar edema, even though this varies wildly in clinical practice. In practice, it is challenging to quantify pulmonary edema at a more granular level. Comparisons between images are easier and more reproducible. Future work could leverage pairs of images to quantify edema on a continuous scale.

The diagnosis of pulmonary edema is often challenging due to the possibility of other competing diagnoses that have overlapping radiographic findings. For example, multifocal pneumonia can be confused with alveolar pulmonary edema, and chronic interstitial edema can be misinterpreted as interstitial pulmonary edema. In order to minimize this bias, we restrict our labeled data to a cohort of patients diagnosed with CHF. In this work, we purposely ignore image findings such as cardiomegaly and pleural effusions that are correlated with pulmonary edema and often used by radiologists when making the diagnosis. In future work, we plan to leverage multi-task training to jointly learn these associated features. By incorporating multiple image observations in the model training, an algorithm would approximate the clinical gestalt that a radiologist has when considering the etiology of pulmonary opacities. By separating the features of pulmonary edema from features that are associated with CHF, however, our model is not biased against detecting non-cardiogenic pulmonary edema.

Lastly, we compare our results only to the chest radiograph rather than some other reference standard of pulmonary edema. In clinical practice, the chest radiograph is usually considered the reference standard to measure pulmonary edema. Pulmonary capillary wedge pressure might be more accurate, but is extremely invasive, and performed only on a small fraction of patients, so would be impractical to be used as a reference standard.

Accurate grading of pulmonary edema on chest radiographs is a clinically important task. The models developed in this study were capable of classifying edema grades on chest radiographs.

## 5. Acknowledgements

The authors thank Alistair Johnson, James L. Smith, Stanley Y. Kim, Amalie C. Thavikulwat for helping with the data curation.

Research reported in this publication was supported by NIH NIBIB NAC P41EB015902, Philips, Wistron, MIT Lincoln Laboratory, and MIT Deshpande Center.

# References

1.  Mahdyoon H, Klein R, Eyler W, Lakier JB, Chakko SC, Gheorghiade M. Radiographic pulmonary congestion in end-stage congestive heart failure. Am. J. Cardiol. 1989 Mar 1;63(9):625–7.

2.  Logue RB, Rogers JV, Gay BB. Subtle roentgenographic signs of left heart failure. Am. Heart J. 1963 Apr;65(4):464–73.

3.  Harrison MO, Conte PJ, Heitzman ER. Radiological detection of clinically occult cardiac failure following myocardial infarctionl. Br. J. Radiol. 1971 Apr;44(520):265–72.

4.  Milne EN. Correlation of physiologic findings with chest roentgenology. Radiol. Clin. North Am. 1973 Apr;11(1):17–47.

5.  Van de Water JM, Sheh JM, O'Connor NE, Miller IT, Milne EN. Pulmonary extravascular water volume: measurement and significance in critically ill patients. J. Trauma. 1970 Jun;10(6):440–9.

6.  Noble WH, Sieniewicz DJ. Radiological changes in controlled hypervolaemic pulmonary oedema in dogs. Canad. Anaesth. Soc. J. 1975 Mar;22(2):171–85.

7.  Snashall PD, Keyes SJ, Morgan BM, McAnulty RJ, Mitchell-Heggs PF, McIvor JM, et al. The radiographic detection of acute pulmonary oedema. A comparison of radiographic appearances, densitometry and lung water in dogs. Br. J. Radiol. 1981 Apr;54(640):277–88.

8.  Gheorghiade M, Follath F, Ponikowski P, Barsuk JH, Blair JEA, Cleland JG, et al. Assessing and grading congestion in acute heart failure: a scientific statement from the acute heart failure committee of the heart failure association of the European Society of Cardiology and endorsed by the European Society of Intensive Care Medicine. Eur. J. Heart Fail. 2010 May;12(5):423–33.

9.  Hunt SA, Abraham WT, Chin MH, Feldman AM, Francis GS, Ganiats TG, et al. 2009 Focused update incorporated into the ACC/AHA 2005 Guidelines for the Diagnosis and Management of Heart Failure in Adults A Report of the American College of Cardiology Foundation/American Heart Association Task Force on Practice Guidelines Developed in Collaboration With the International Society for Heart and Lung Transplantation. J. Am. Coll. Cardiol. 2009 Apr 14;53(15):e1–90.

10. Adams KF, Fonarow GC, Emerman CL, LeJemtel TH, Costanzo MR, Abraham WT, et al. Characteristics and outcomes of patients hospitalized for heart failure in the United States: rationale, design, and preliminary observations from the first 100,000 cases in the Acute Decompensated Heart Failure National Registry (ADHERE). Am. Heart J. 2005 Feb;149(2):209–16.

11. Deng J, Dong W, Socher R, Li L-J, Li K, Fei-Fei L. ImageNet: A large-scale hierarchical image database. 2009 IEEE Conference on Computer Vision and Pattern Recognition. IEEE; 2009. p. 248–55.

12. Johnson AEW, Pollard TJ, Berkowitz SJ, Greenbaum NR, Lungren MP, Deng C, et al.

MIMIC-CXR: A large publicly available database of labeled chest radiographs. CoRR. 2019;abs/1901.07042.

13. Johnson AEW. The MIMIC-CXR Database. physionet.org. 2019;

14. Goldberger AL, Amaral LA, Glass L, Hausdorff JM, Ivanov PC, Mark RG, et al. PhysioBank, PhysioToolkit, and PhysioNet: components of a new research resource for complex physiologic signals. Circulation. 2000 Jun 13;101(23):E215-20.

15. Johnson AEW, Pollard TJ, Berkowitz SJ, Greenbaum NR, Lungren MP, Deng C-Y, et al. MIMIC-CXR, a de-identified publicly available database of chest radiographs with free-text reports. Sci. Data. 2019 Dec 12;6(1):317.

16. Liao R, Rubin J, Lam G, Berkowitz S, Dalal S, Wells W, Horng S, Golland P. Semi-supervised learning for quantification of pulmonary edema in chest x-ray images. arXiv preprint arXiv:1902.10785. 2019 Feb 27.

17. Huang G, Liu Z, Maaten L van der, Weinberger KQ. Densely connected convolutional networks. 2017 IEEE Conference on Computer Vision and Pattern Recognition (CVPR). IEEE; 2017. p. 2261–9.

18. Wang X, Schwab E, Rubin J, Klassen P, Liao R, Berkowitz S, et al. Pulmonary Edema Severity Estimation in Chest Radiographs Using Deep Learning. N/A; 2019.

19. Selvaraju RR, Cogswell M, Das A, Vedantam R, Parikh D, Batra D. Grad-CAM: Visual explanations from deep networks via gradient-based localization. Proceedings of 2017 IEEE International Conference on Computer Vision (ICCV). IEEE; 2017. p. 618–26.

20. Wang X, Peng Y, Lu L, Lu Z, Bagheri M, Summers RM. ChestX-Ray8: Hospital-Scale Chest X-Ray Database and Benchmarks on Weakly-Supervised Classification and Localization of Common Thorax Diseases. 2017 IEEE Conference on Computer Vision and Pattern Recognition (CVPR). IEEE; 2017. p. 3462–71.

21. Dunnmon JA, Yi D, Langlotz CP, Ré C, Rubin DL, Lungren MP. Assessment of convolutional neural networks for automated classification of chest radiographs. Radiology. 2019;290(2):537–44.

22. Rajpurkar P, Irvin J, Zhu K, Yang B, Mehta H, Duan T, et al. CheXNet: Radiologist-Level Pneumonia Detection on Chest X-Rays with Deep Learning. arXiv. 2017 Nov 14;

23. Majkowska A, Mittal S, Steiner DF, Reicher JJ, McKinney SM, Duggan GE, et al. Chest Radiograph Interpretation with Deep Learning Models: Assessment with Radiologist-adjudicated Reference Standards and Population-adjusted Evaluation. Radiology. 2019 Dec 3;191293.

24. Seah JCY, Tang JSN, Kitchen A, Gaillard F, Dixon AF. Chest radiographs in congestive heart failure: visualizing neural network learning. Radiology. 2019 Feb;290(2):514–22.

25. Hammon M, Dankerl P, Voit-Höhne HL, Sandmair M, Kammerer FJ, Uder M, et al. Improving diagnostic accuracy in assessing pulmonary edema on bedside chest radiographs using a standardized scoring approach. BMC Anesthesiol. 2014 Oct 18;14:94.

26. Zhao CY, Xu-Wilson M, Gangireddy SR, Horng S. Predicting Disposition Decision, Mortality, and Readmission for Acute Heart Failure Patients in the Emergency Department Using Vital Sign, Laboratory, Echocardiographic, and Other Clinical Data. Circulation. 2018 Nov 6;138(Suppl_1):A14287-.

**Table 1: Validation of Keyword Terms**

| Edema severity | Keyword | Number of reports | Precision | Sensitivity | Specificity |
|---|---|---|---|---|---|
| "Overall" | N/A | 200 | 92% | N/A | N/A |
| None | No pulmonary edema | 24 | 95.83% | 40.35% | 99.41% |
| | No vascular congestion | 18 | 94.44% | 29.82% | 99.41% |
| | No fluid overload | 2 | 100% | 3.51% | 100% |
| | No acute cardiopulmonary process | 13 | 92.31% | 21.05% | 99.41% |
| Vascular congestion | Cephalization | 24 | 75% | 33.96% | 96.55% |
| | Mild pulmonary vascular congestion | 24 | 91.67% | 41.51% | 98.85% |
| | Mild hilar engorgement | 2 | 100% | 3.77% | 100% |
| | Mild vascular plethora | 8 | 100% | 15.09% | 100% |
| Interstitial edema | Interstitial opacities | 15 | 93.33% | 20.90% | 99.38% |
| | Kerley | 19 | 100% | 28.36% | 100% |
| | Interstitial edema | 20 | 100% | 29.85% | 100% |
| | Interstitial thickening | 8 | 75% | 8.96% | 98.75% |
| Alveolar edema | Alveolar infiltrates | 16 | 100% | 32.00% | 100% |
| | Severe pulmonary edema | 33 | 90.91% | 60.00% | 98.87% |
| | Perihilar infiltrates | 1 | 100% | 2.00% | 100% |
| | hilar infiltrates | 1 | 100% | 2.00% | 100% |

**Note.** The total number of reports from all the keywords is more than 200 because some reports have more than one keyword. The low sensitivity and high specificity of each keyword indicate

that no single keyword can represent the entire severity level but every keyword is specific to the severity level that it is supposed to belong to.

**Table 2: Distribution of Severity Labels across Folds and Test Set**

| Fold | 0 | 1 | 2 | 3 | Total Images |
|---|---|---|---|---|---|
| | None | Vascular congestion | Interstitial edema | Alveolar edema | |
| A. Unlabeled | | | | | |
| ($n$ = 63 149) | -- | -- | -- | -- | 229,519 |
| B. Labeled-regular expressions (cross validation) | | | | | |
| Fold 1 ($n$ = 254) | 260 | 130 | 189 | 27 | 606 |
| Fold 2 ($n$ = 253) | 296 | 150 | 215 | 31 | 692 |
| Fold 3 ($n$ = 253) | 269 | 130 | 236 | 26 | 661 |
| Fold 4 ($n$ = 253) | 292 | 153 | 194 | 38 | 677 |
| Fold 5 ($n$ = 253) | 302 | 153 | 237 | 26 | 718 |
| Sub-total ($n$ = 1266) | 1,419 (42.13%) | 716 (21.35%) | 1071 (31.93%) | 148 (4.41%) | 3354 (100%) |
| C. Labeled-manual (test) | | | | | |
| ($n$ = 123) | 61 (43.26%) | 44 (31.21%) | 20 (14.18%) | 16 (11.35%) | 141 (100%) |

**Note.** Regular expressions indicate the retrieval of text from the original radiology reports within the database.

**Table 3: Patient Demographics and Characteristics**

|  | CHF (n = 1916) | | | Non-CHF (n = 62 665) | P-value |
|---|---|---|---|---|---|
|  | Labelled (n = 1266) | Unlabeled (n = 650) | Total (n = 1916) | | |
| Age – years (95% CI) | 73 (72.0-74.1) | 75.8 (75.2-76.4) | 75.1 (74.5-75.6) | 51.0 (50.9-51.1) | < .001 |
| Women (%) | 51.8% | 51.3% | 51.4% | 54.6% | .001 |
| Disposition | | | | | < .001 |
| Admit (%) | 91.5% | 93.6% | 92.8% | 35.6% | |
| Discharge (%) | 8.2% | 5.9% | 6.5% | 59.6% | |
| AMA (%) | 0.0% | 0.2% | 0.2% | 0.3% | |
| Cardiac catheterization | 0.0% | 0.1% | 0.0% | 0.1% | |
| Eloped | 0.0% | 0.0% | 0.0% | 1.1% | |
| Expired | 0.0% | 0.2% | 0.1% | 0.1% | |
| Labor & Delivery | 0.0% | 0.0% | 0.0% | 0.0% | |
| LWBS | 0.2% | 0.0% | 0.0% | 1.1% | |
| OR | 0.2% | 0.1% | 0.1% | 0.7% | |
| Transfer | 0.0% | 0.0% | 0.2% | 1.4% | |
| Number of CXRs | | | 9 (1-153) | 3 (1-174) | |
| Interval, days | | | 7.09 (0.13-1545) | | |

**Note.** Number of chest radiographs (CXRs) per patient and the interval time between two chest radiographs are shown as median (range). Interval indicates the interval between two consecutive chest radiographs from the same patient.

**Table 4: AUC from the Semi-supervised Model and the Pre-trained Supervised Learning Model on the Test Set**

| Comparison | Semi-supervised | Pre-trained supervised | P value* |
|---|---|---|---|
| 0 vs 1 | 0.79 | 0.66 | .02 |
| 0 vs 2 | 0.88 | 0.81 | .29 |
| 0 vs 3 | 0.99 | 0.87 | .003 |
| 1 vs 2 | 0.69 | 0.73 | .58 |
| 1 vs 3 | 0.93 | 0.82 | .07 |
| 2 vs 3 | 0.88 | 0.63 | .01 |
| 0 vs 1, 2, 3 | 0.85 | 0.74 | .008 |
| 0, 1 vs 2, 3 | 0.88 | 0.81 | .15 |
| 0, 1, 2 vs 3 | 0.96 | 0.82 | .002 |

*Significance testing between the semi-supervised model and the pre-trained supervised model area under the curve using DeLong's method (p-value of the hypothesis that they have the same performance). In order to account for multiple comparisons, a Bonferroni correction was used where a P value below .005 indicates a significant difference ($\alpha = 0.05 / 9 = 0.005$). All the results are based on the predictions of the test set.

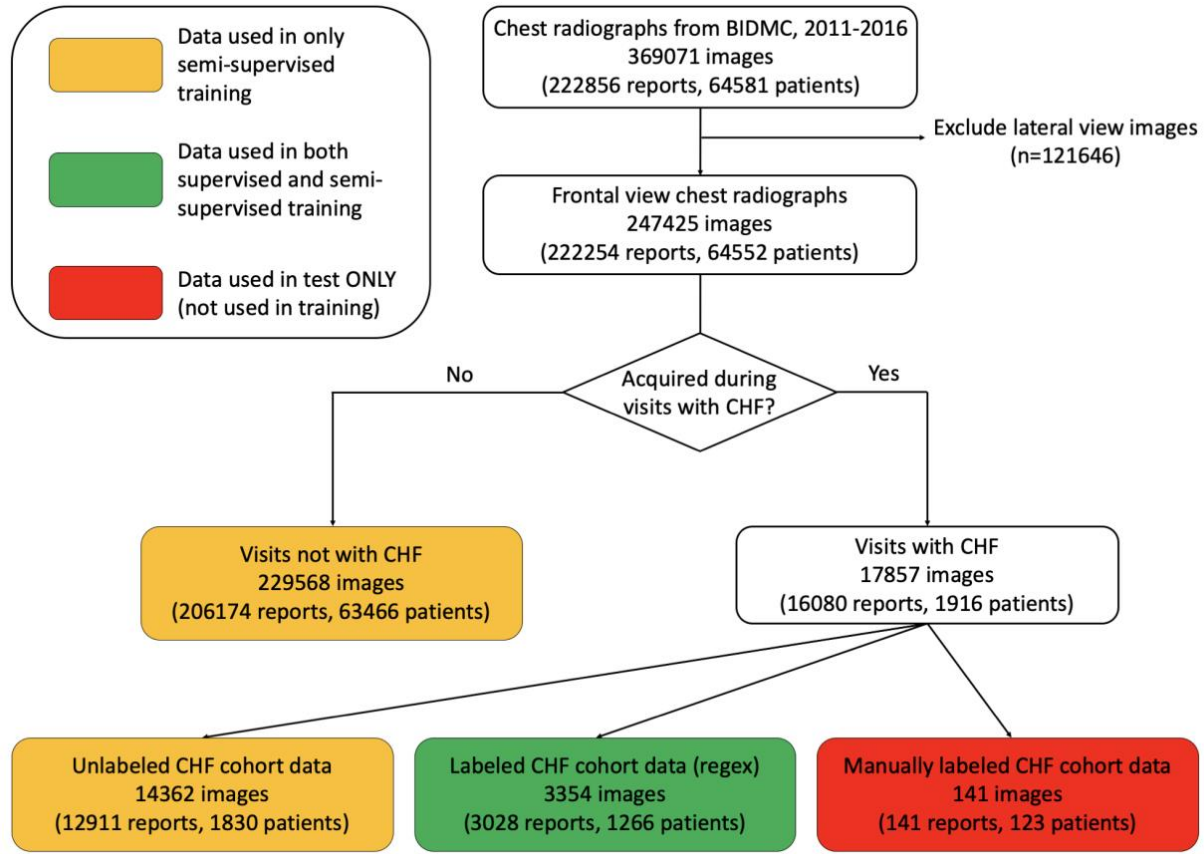

**Figure 1:** Cohort selection flowchart. 369,071 chest radiographs and their associated radiology reports from 62,665 patients were collected. Images for this study were limited to frontal view radiographs (247,425). Of the 247,425 frontal view radiographs, 17,857 images were acquired during visits with a diagnosis consistent with congestive heart failure. In the CHF cohort, we were able to label 3,028 radiology reports and thus 3,354 frontal view radiographs from 1266 patients, using regular expressions on the reports. We also curated a test set of 141 radiographs that were manually labeled by radiologists (from the 650 unlabeled radiographs from patients with CHF).

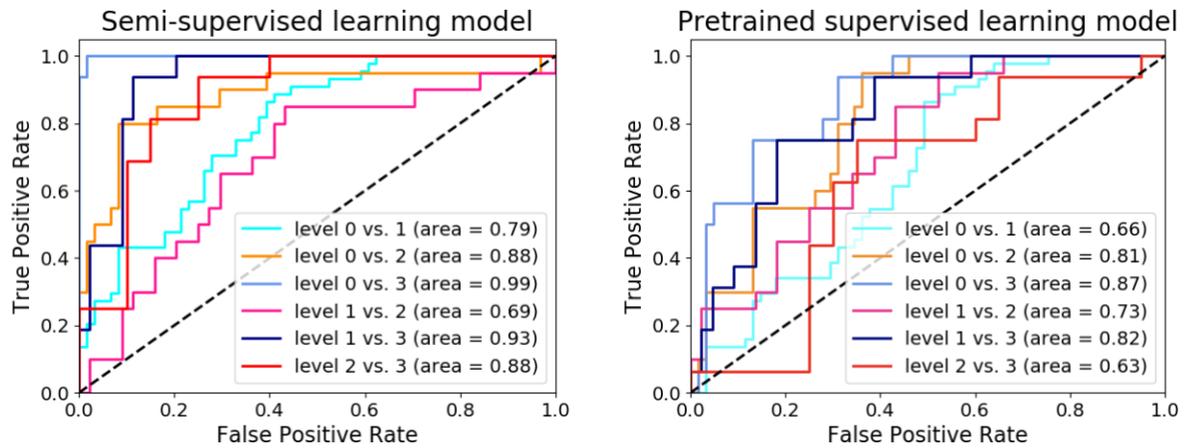

a.

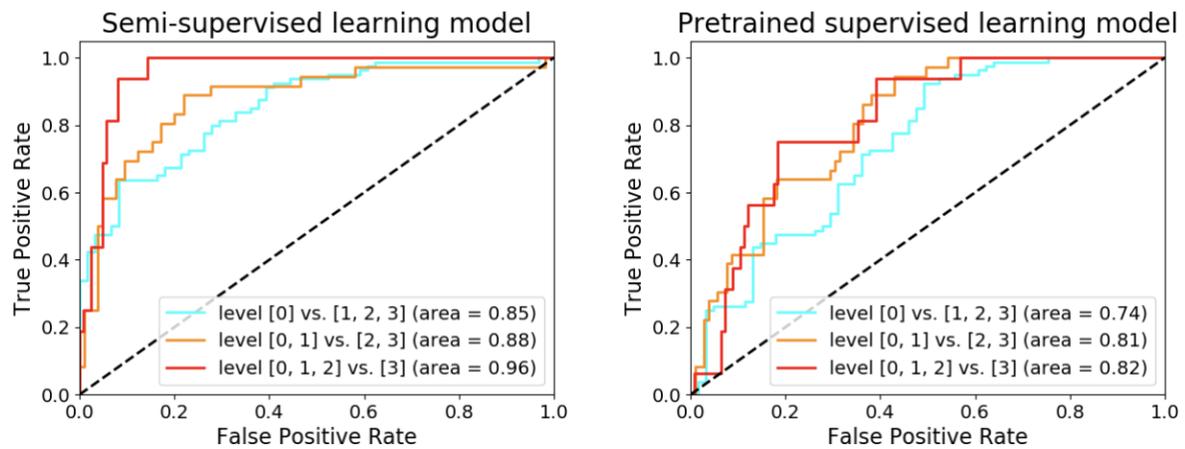

b.

**Figure 2:** Receiver operating characteristic (ROC) curves of the semi-supervised learning model and the pre-trained supervised learning model. All the curves are based on the predictions of the test set. **(a)** ROC curves for six pairwise comparisons. **(b)** ROC curves for three dichotomized severity comparisons. All the curves are based on the predictions of the test set.

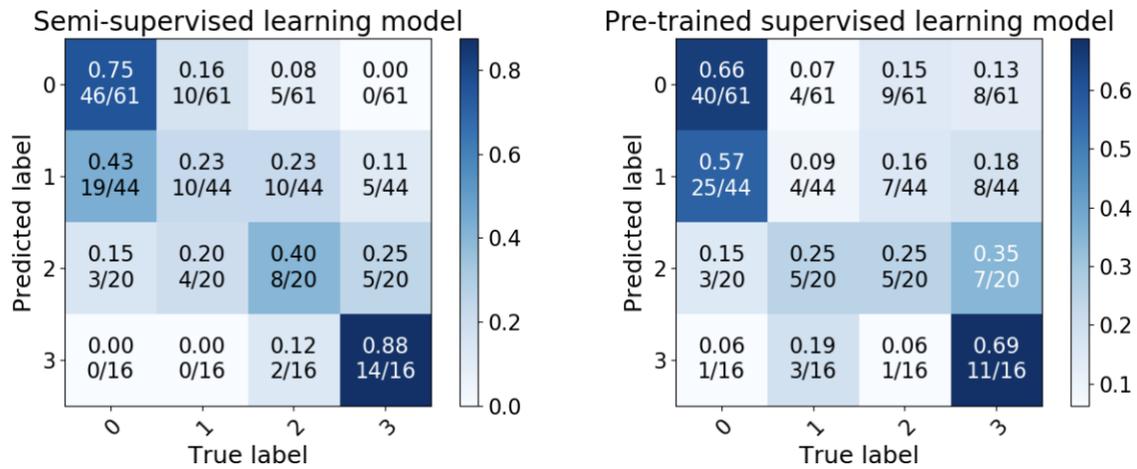

**Figure 3:** Confusion matrices from the semi-supervised learning model and the pre-trained supervised learning model. The denominator of each fraction number is the number of images that the algorithm predicts of the corresponding row, and the numerator is the number of images that belongs to the corresponding column. The quadratic-weighted Kappa values of the semi-supervised learning model and the pre-trained supervised learning model are 0.70 and 0.41. All the results are based on the predictions of the test set.

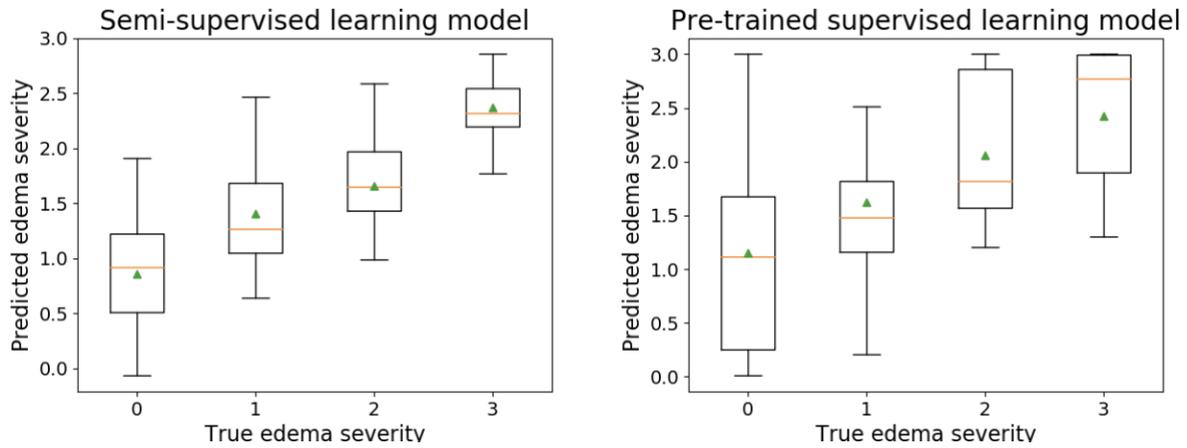

**Figure 4:** Predicted edema severity scores versus true edema severity labels from the semi-supervised learning model and the pre-trained supervised learning model. The box extends from the lower to upper quartile values of the distribution, with the orange line at the median and the green triangle at the mean. The whiskers extend from the box to show the range of the data. All the results are based on the predictions of the test set.

Semi-supervised learning model | Pre-trained supervised learning model

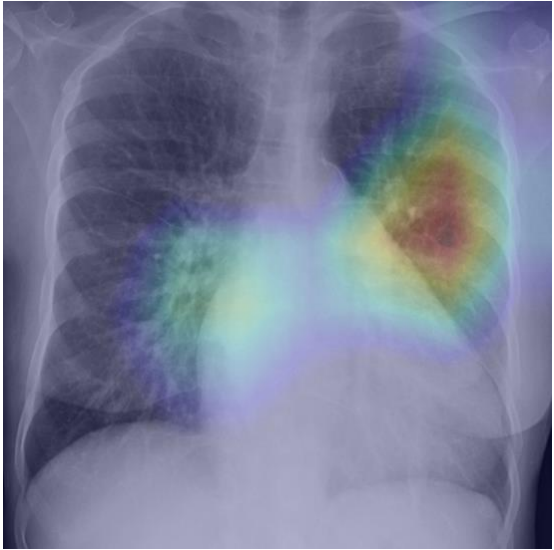 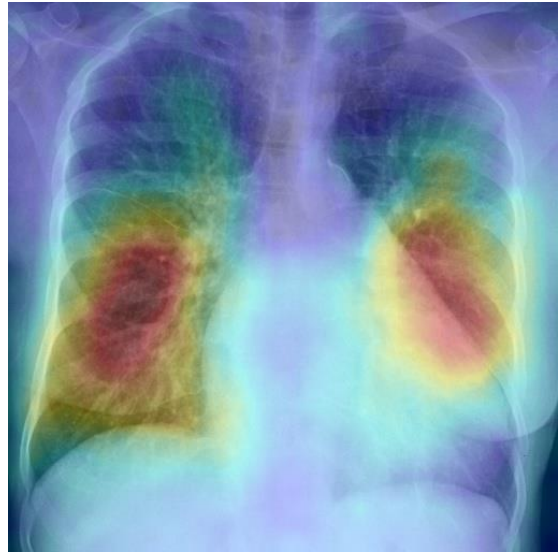

a.

Semi-supervised learning model | Pre-trained supervised learning model

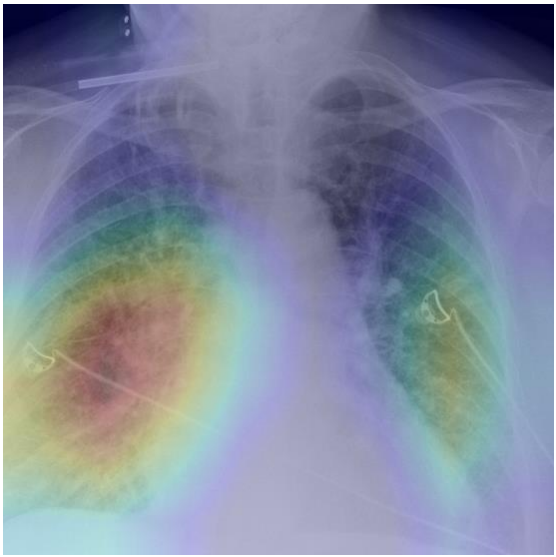 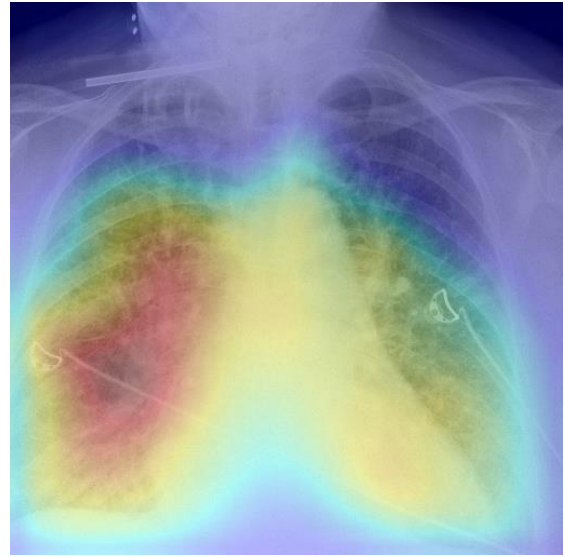

b.

**Figure 5:** Grad-CAM heatmaps that highlight important regions for the model prediction. **(a)** A sample radiograph that is labeled as "vascular congestion" (level 1). **(b)** A sample radiograph that is labeled as "alveolar edema" (level 3).

# Appendix E1

**Background Information**

Accurate monitoring of pulmonary edema is essential when competing clinical priorities complicate clinical management. For example, a CHF patient with a severe infection causing septic shock may have pulmonary edema driven both by volume overload in heart failure and increased capillary permeability. This patient will likely be intravascularly depleted from their septic shock, but also total body volume overloaded, leading to pulmonary edema. The patient simultaneously needs both more fluid to optimize their hemodynamic function and less fluid to optimize their respiratory function. Often referred to as the ebb and flow of sepsis, patients need judicious fluid resuscitation early in their clinical course, and evacuation of fluid through diuresis later in their course (1,2). The accurate assessment of pulmonary edema is critical to maintaining this delicate fluid balance.

Decompensated CHF patients have heterogeneous responses to treatment (3), and that response is highly predictive of clinical trajectory. However, this response to treatment is poorly documented in the medical record, limiting the ability of researchers to discover important relationships between treatments and effects. Other surrogates for response to treatment such as urine output, total body fluid balance, and daily weights have been suggested, but are often not accurately and consistently measured.

Although improvement in dyspnea correlates with radiographic improvement, critically ill patients cannot provide this information and subjective information is not well quantified. The automatic and quantitative assessment for pulmonary edema severity will enable clinicians to make better treatment plans based on prior patient responses and will also enable clinical research studies that require quantitative phenotyping of patient status (4).

**Reference standard image labelling**

We performed a modified Delphi consensus process to develop a gold standard image label. We had 3 senior radiology residents and 1 attending radiologist manually label a set of 141 frontal view chest radiographs from 123 patients. The three residents labeled the images independently. If the three residents had exactly the same pulmonary edema severity of an image, then a consensus label is assigned. If only two out of the three residents agreed on the edema severity, then an attending radiologist reviewer was added. If a majority of the reviewers (three out of four) now agreed, then a consensus label is assigned. If no consensus was reached, then the four radiologists discussed their interpretations in a round-robin process, and then again voted anonymously on their edema severity levels. If a majority of the votes was reached, then a consensus label is assigned. If no consensus was reached, then another round-robin discussion is performed with another anonymous vote. This process is then repeated one additional time, and if no consensus is reached, then the image is labelled as no consensus. The flowchart of the consensus process is shown in the **Figure E2**.

**References**

1. Jaehne AK, Rivers EP. Early liberal fluid therapy for sepsis patients is not harmful: hydrophobia is unwarranted but drink responsibly. Crit. Care Med. 2016;44(12):2263–9.
2. Malbrain MLNG, Van Regenmortel N, Saugel B, De Tavernier B, Van Gaal P-J, Joannes-Boyau O, et al. Principles of fluid management and stewardship in septic shock: it is time to consider the four D's and the four phases of fluid therapy. Ann. Intensive Care. 2018 May 22;8(1):66.
3. Francis GS, Cogswell R, Thenappan T. The heterogeneity of heart failure: will enhanced phenotyping be necessary for future clinical trial success? J. Am. Coll. Cardiol. 2014 Oct 28;64(17):1775–6.
4. Chakko S, Woska D, Martinez H, de Marchena E, Futterman L, Kessler KM, et al. Clinical, radiographic, and hemodynamic correlations in chronic congestive heart failure: conflicting results may lead to inappropriate care. Am. J. Med. 1991 Mar;90(3):353–9.
5. He K, Zhang X, Ren S, Sun J. Deep residual learning for image recognition. IEEE

Conference on Computer Vision and Pattern Recognition (CVPR). IEEE; 2016. p. 770–8.

6. Huang G, Liu Z, Maaten L van der, Weinberger KQ. Densely connected convolutional networks. 2017 IEEE Conference on Computer Vision and Pattern Recognition (CVPR). IEEE; 2017. p. 2261–9.

7. Deng J, Dong W, Socher R, Li L-J, Li K, Fei-Fei L. ImageNet: A large-scale hierarchical image database. 2009 IEEE Conference on Computer Vision and Pattern Recognition. IEEE; 2009. p. 248–55.

| Pulmonary Edema Severity Level Label | Pathophysiology | Representative Chest Radiograph | Common Radiographic Findings | Pulmonary Capillary Wedge Pressure |
|---|---|---|---|---|
| 0 | None | 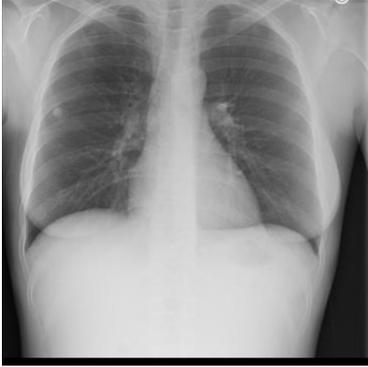 | | |
| 1 | Vascular congestion | 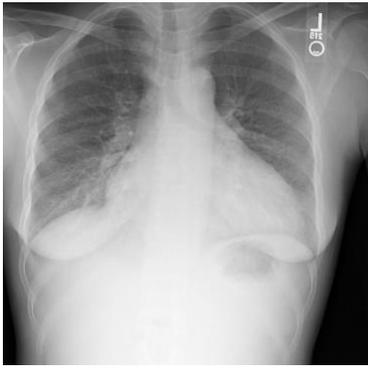 | cephalization, pulmonary vascular congestion, hilar vascular indistinctness | 13-18 mm Hg |
| 2 | Interstitial edema | 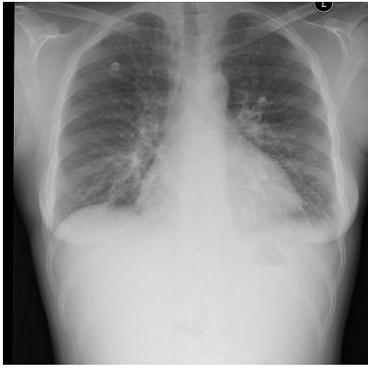 | increased interstitial markings, Kerley B lines, peribronchial cuffing | 18-25 mm Hg |

| 3 | Alveolar edema | 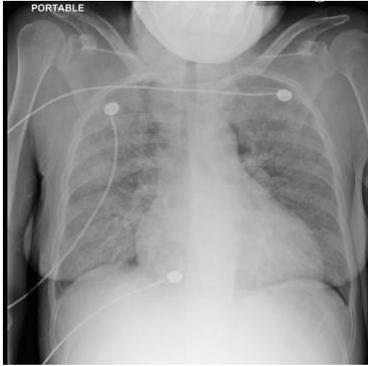 | bilateral, symmetric, airspace opacities radiating centrally from the hila

pleural effusion | >25 mm Hg |

**Figure E1:** Representative images and radiographic findings of each pulmonary edema severity level.

**Modified Delphi Consensus Process**

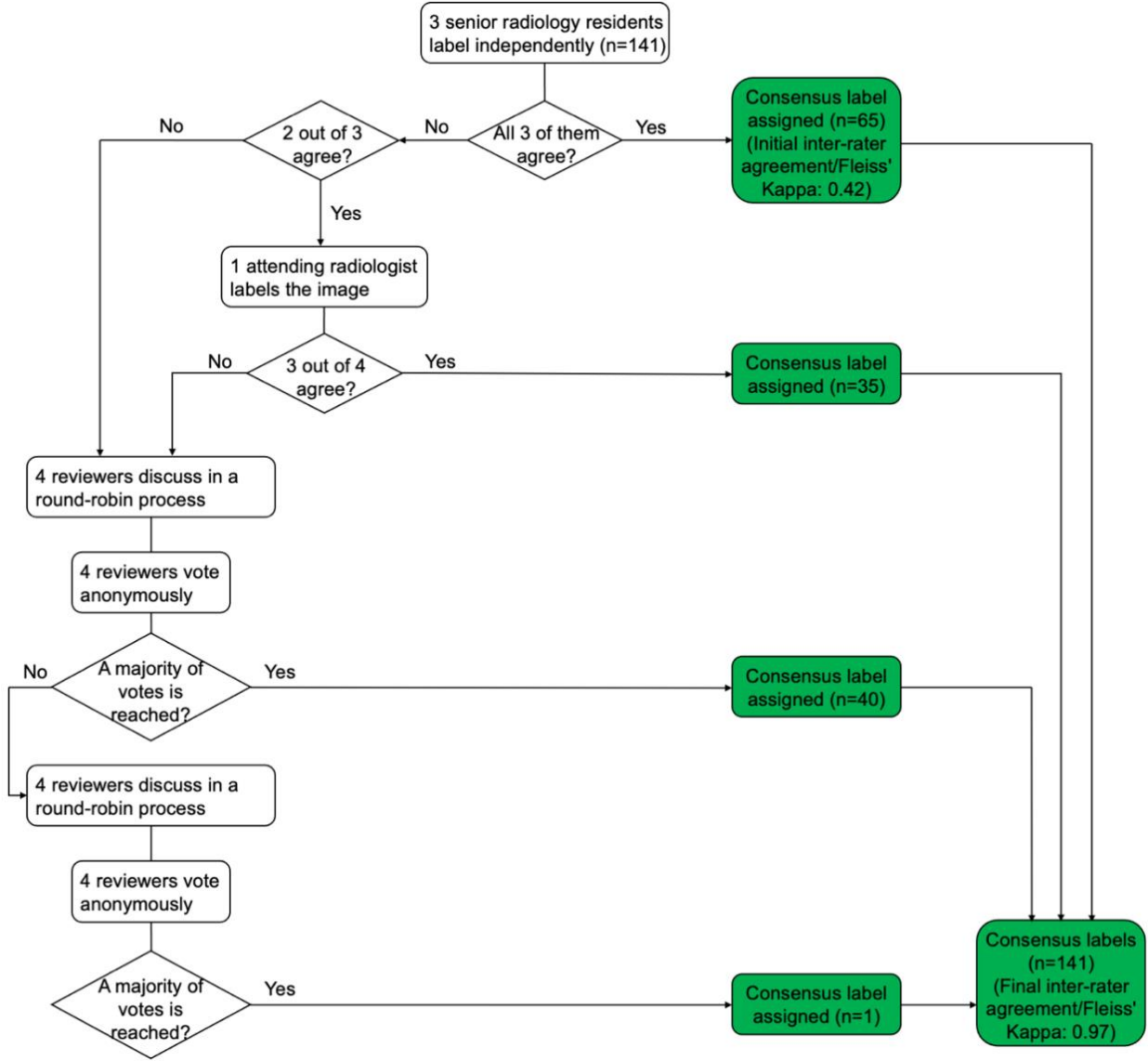

**Figure E2:** The flowchart of our consensus image labeling process. The initial labels independently provided by the 3 senior radiology residents against the final consensus labels have quadratic-weighted Kappa values of 0.83, 0.74, and 0.72. The predictions from the semi-supervised learning model and the pre-trained supervised learning model against the final consensus labels have quadratic-weighted Kappa values of 0.70 and 0.41.

**Chest Radiograph Distributions**

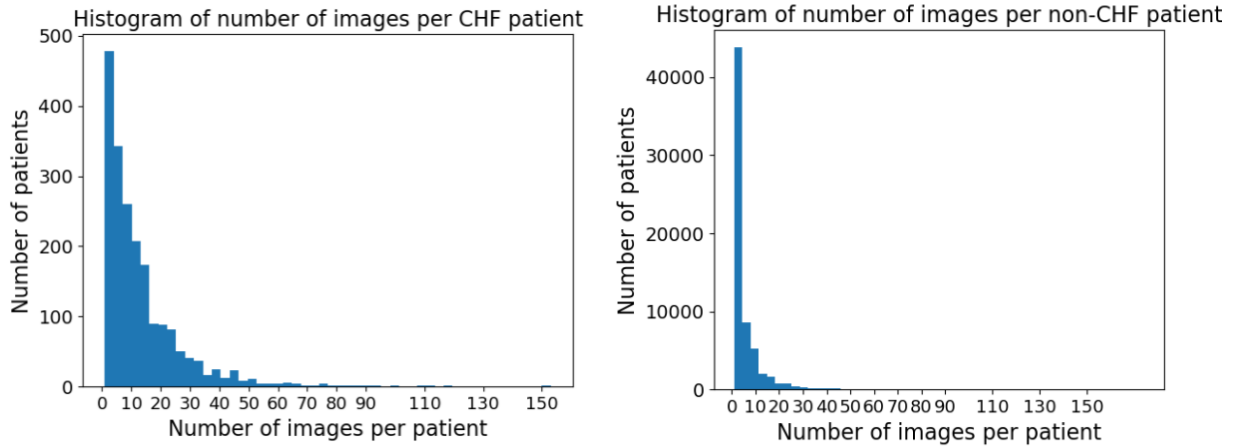

a.

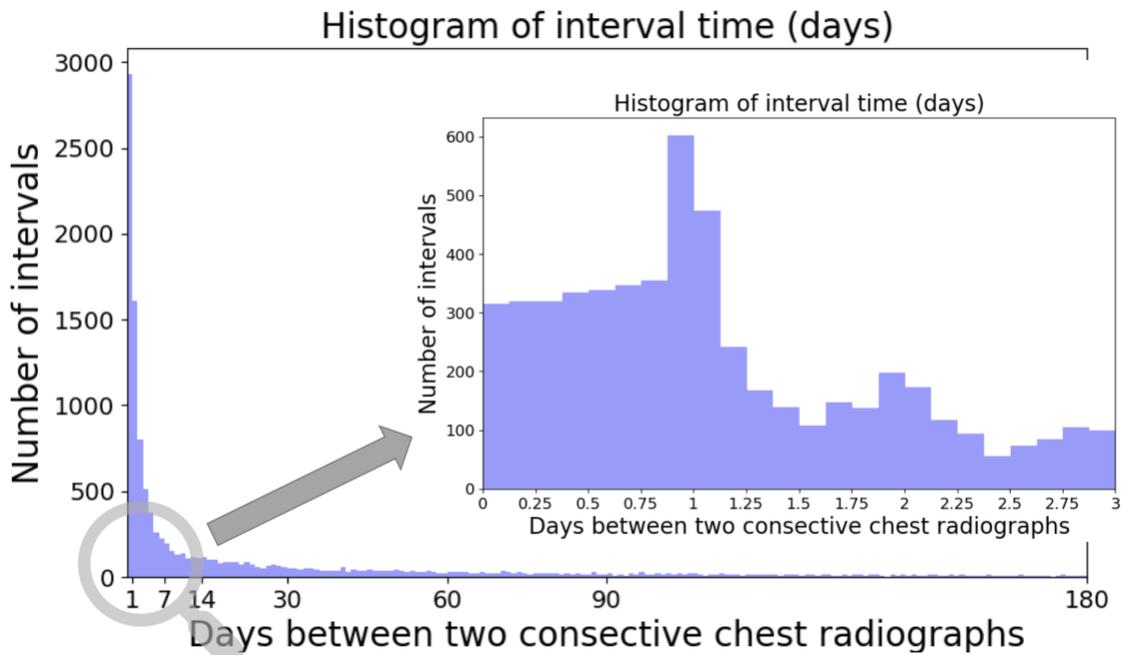

b.

**Figure E3:** Chest radiograph distributions. **(a)** Histograms of the number of images per CHF patient and per non-CHF patient. On average, 13.78 chest radiographs were taken per CHF patient and 5.43 chest radiographs were taken per non-CHF patient in our dataset. The median number of chest radiographs taken per CHF patient is 9 (ranging from 1 to 153) and per non-CHF

patient is 3 (ranging from 1 to 174). **(b)** Distributions of time intervals between serial chest radiographs in CHF cohort. The x-axis is in log scale. The mean interval time between each two consecutive chest radiographs of the same CHF patient is 71.34 days. The median interval time between each two consecutive chest radiographs of the same CHF patient is 7.09 days (ranging from 180 minutes to 1545.84 days). 21.53% of the interval times for CHF patients are within 1 day and 66.08% are within 30 days.

**Supplemental Table 1: Semi-supervised Model with a Varied Number of Unlabeled CXRs**

| Number of unlabeled CXRs | 0 | 66000 | 110000 | 154000 | 198000 | 220000 | 233284 |
|---|---|---|---|---|---|---|---|
| Percentage of total unlabeled CXRs | 0% | 28.29% | 47.15% | 66.01% | 84.88% | 94.31% | 100% |
| Average AUC | 0.66 | 0.80 | 0.80 | 0.82 | 0.82 | 0.84 | 0.87 |

**Note.** All the results are based on the predictions of the test set.

**Cross Validation Results**

**Supplemental Table 2: AUC from the Semi-supervised Model and the Pre-trained Supervised Learning Model on Cross-Validation**

| Comparison | Semi-supervised | Pre-trained supervised | *P* value* |
|---|---|---|---|
| 0 vs 1 | 0.77 (±0.03) | 0.76 (±0.02) | .65 |
| 0 vs 2 | 0.82 (±0.01) | 0.83 (±0.02) | .70 |
| 0 vs 3 | 0.97 (±0.02) | 0.96 (±0.02) | .14 |
| 1 vs 2 | 0.59 (±0.03) | 0.63 (±0.04) | .81 |
| 1 vs 3 | 0.92 (±0.03) | 0.88 (±0.05) | .33 |
| 2 vs 3 | 0.88 (±0.04) | 0.81 (±0.07) | .02 |
| 0 vs 1, 2, 3 | 0.81 (±0.01) | 0.81 (±0.02) | .89 |
| 0, 1 vs 2, 3 | 0.77 (±0.01) | 0.78 (±0.02) | .31 |
| 0, 1, 2 vs 3 | 0.93 (±0.03) | 0.89 (±0.04) | .03 |

**Note.** The average area under the curve and its standard deviation of the five folds are reported in the table. Ss: AUC from the semi-supervised learning model. Ps: AUC from the pre-trained supervised learning model. Delong: significance testing between ss and ps AUC using DeLong's method (p-value of the hypothesis that they have the same performance). All the results in this table come from the 5-fold cross-validation.

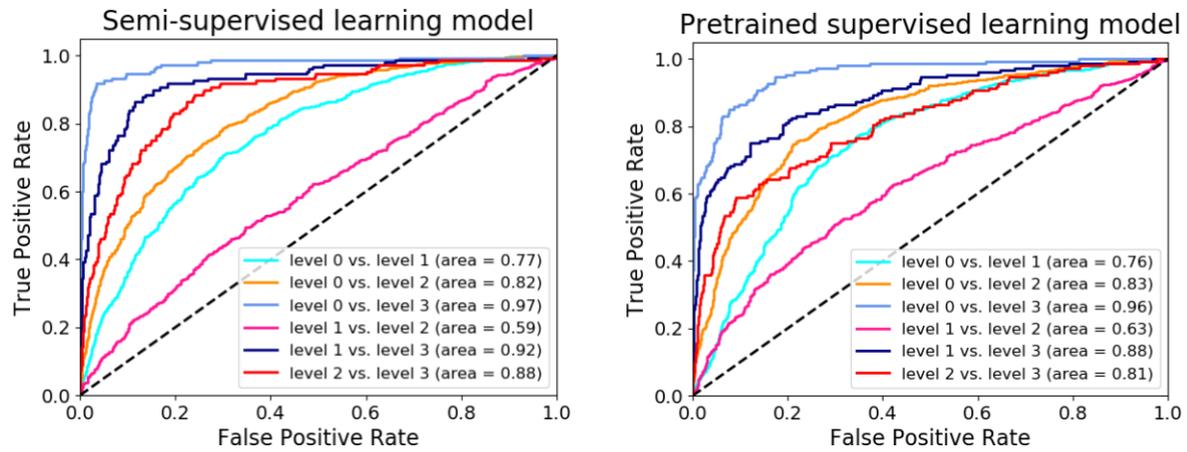

a.

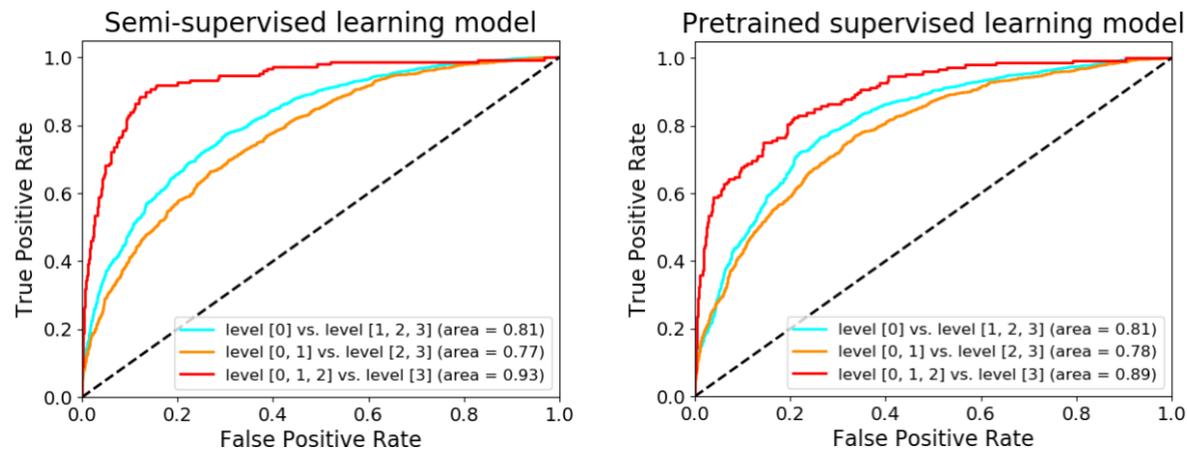

b.

**Figure E4:** Receiver operating characteristic (ROC) curves of the semi-supervised learning model and the pre-trained supervised learning model. All the curves are based on the predictions of the five folds from cross-validation. **(a)** ROC curves for 6 pairwise comparisons. **(b)** ROC curves for 3 dichotomized severity comparisons. All the curves in this figure are based on the predictions from the 5-fold cross-validation.

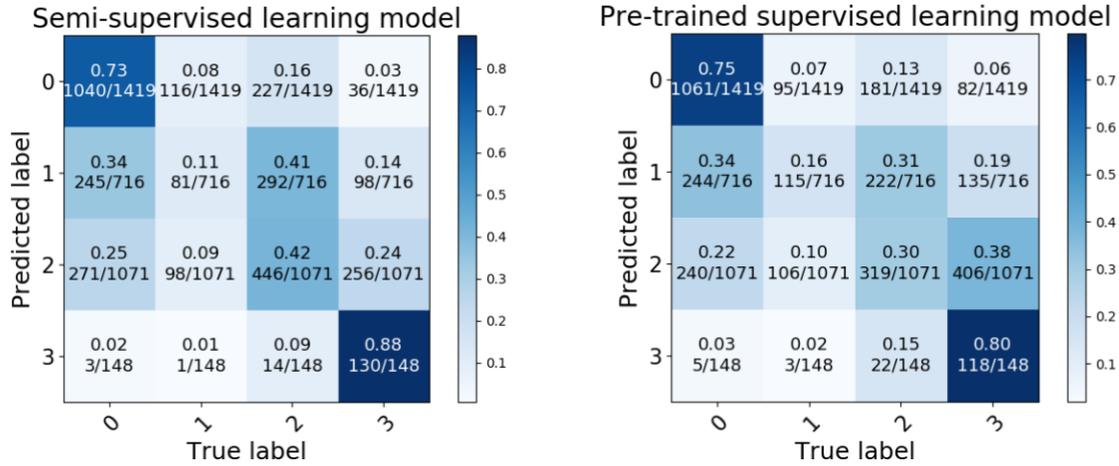

**Figure E5:** Confusion matrices from the semi-supervised learning model and the pre-trained supervised learning model. We show counts for each cell and row percentages. All the results in this figure are based on the predictions from the 5-fold cross-validation.

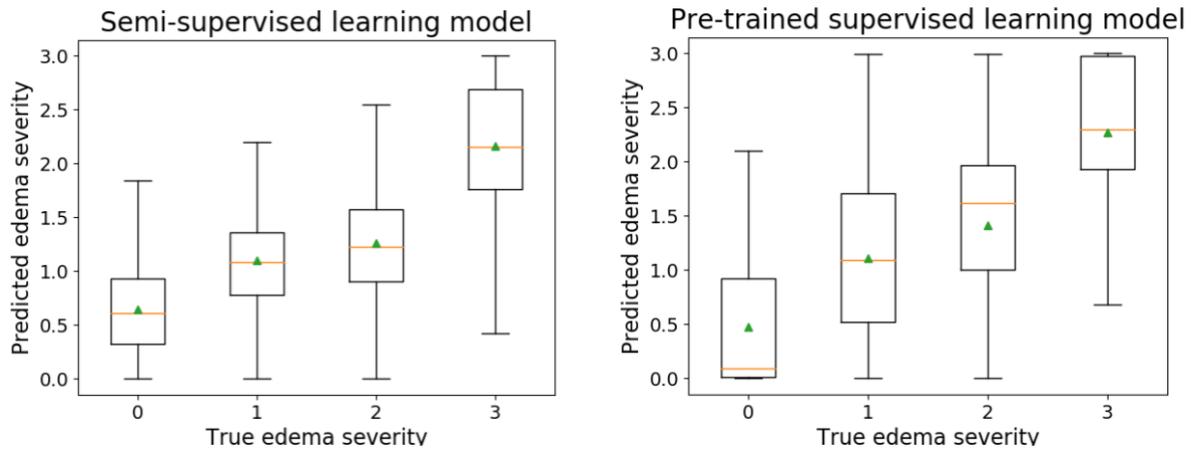

**Figure E6:** Predicted edema severity scores versus true edema severity labels from the semi-supervised learning model and the pre-trained supervised learning model. The box extends from the lower to upper quartile values of the distribution, with the orange line at the median and the green triangle at the mean. The whiskers extend from the box to show the range of the data. Both plots in this figure are based on the predictions from the 5-fold cross-validation.

**Critical Analysis of Results**

**Supplemental Table 3: Failure Modes of the Semi-supervised and Pre-trained Models**

|  | Failure mode | Semi-supervised model | | Pre-trained model | |
| --- | --- | --- | --- | --- | --- |
|  |  | Number of images | Percentage in the test set | Number of images | Percentage in the test set |
| Disagree | In-between | 38 | 26.95% | 38 | 26.95% |
|  | Alternate pathology | 9 | 6.38% | 18 | 12.77% |
|  | Low lung volumes | 3 | 2.13% | 3 | 2.13% |
|  | Poor exposure | 3 | 2.13% | 7 | 4.96% |
|  | Patient positioning | 1 | 0.71% | 1 | 0.71% |
|  | External devices | 1 | 0.71% | 0 | 0.00% |
|  | Unknown | 8 | 5.67% | 14 | 9.93% |
| Agree |  | 78 | 55.32% | 60 | 42.55% |
| Total |  | 141 |  | 141 |  |

**Note.** Our labeling schema attempts to categorize pulmonary edema into 4 discrete levels whereas in reality, it exists on a continuous scale as we have discussed in the main manuscript. The failure mode of "in between" indicates that the true degree of pulmonary edema is likely in between two of our discrete categories, therefore the discrepancy between label and algorithm output is not unreasonable.